\documentstyle[epsfig]{mn}
\newif\ifAMStwofonts
\AMStwofontstrue

\def\etal{{et al.}}
\def\sta{St\"ackel}
\def\stapo{St\"ackel potential}
\def\stapos{St\"ackel potentials}
\def\df{distribution function}

\ifnfsstwo
  \DeclareMathAlphabet{\mathfrak}{U}{euf}{m}{n}
  \SetMathAlphabet\mathfrak{bold}{U}{euf}{b}{n}
  \DeclareMathAlphabet{\mathbb}{U}{msb}{m}{n}
  \SetMathAlphabet\mathbb{bold}{U}{msb}{m}{n}
  \DeclareMathAlphabet{\mathbfit}{OT1}{cmr}{bx}{it}
  \SetMathAlphabet\mathbfit{bold}{OT1}{cmr}{bx}{it}
  \DeclareMathAlphabet{\mathbfss}{OT1}{cmss}{bx}{n}
  \SetMathAlphabet\mathbfss{bold}{OT1}{cmss}{bx}{n}
  \ifAMStwofonts
    \ifCUPmtlplainloaded \else
      \DeclareSymbolFont{UPM}{U}{eur}{m}{n}
      \SetSymbolFont{UPM}{bold}{U}{eur}{b}{n}
      \DeclareSymbolFont{AMSa}{U}{msa}{m}{n}
      \DeclareMathSymbol{\upi}{0}{UPM}{"19}
      \DeclareMathSymbol{\umu}{0}{UPM}{"16}
      \DeclareMathSymbol{\upartial}{0}{UPM}{"40}
      \DeclareMathSymbol{\leqslant}{3}{AMSa}{"36}
      \DeclareMathSymbol{\geqslant}{3}{AMSa}{"3E}

       \let\le=\leqslant
       \let\ge=\geqslant
    \fi
  \fi
\fi 

\newcounter{parentequation}

\title[Approximate third integrals for axisymmetric potentials using local 
St\"ackel 
fits]%
{Approximate third integrals for axisymmetric potentials using local St\"ackel 
fits}

\author[De Bruyne, Leeuwin, \& Dejonghe]%
{De Bruyne V., Leeuwin F. and Dejonghe H.\\
Astronomical Observatory, Gent University, Krijgslaan 281 S9, B-9000 Gent, 
Belgium\\
} 

\begin{document}

\maketitle
\begin{abstract} 
We use a set of St\"ackel potentials to obtain a local approximation
for an effective third integral in axisymmetric systems.  We present a
study on the feasibility and effectiveness of this approach.  We have
applied it to three trial potentials of various flattenings,
corresponding to nearly ellipsoidal, disky and boxy density isophotes.
In all three cases, a good fit to the potential requires only a small
set of \sta\ potentials, and the associated \sta\ third integral
provides a very satisfactory, yet analytically simple, approximation
to the trial potentials effective third integral.
\end{abstract}

\section{Introduction}

It is well-known that axisymmetric potentials do not in general allow
a global and isolating third integral in addition to the energy and
the component of the angular momentum along the symmetry axis.
However, for many axisymmetric potentials, numerical experience shows
that the majority of orbits are constrained by an effective third
integral over a large number of orbital periods.  As appears from
modeling observed galaxies, a third integral is often required to
describe their dynamical structure ({\it e.g.} Binney \etal 1990;
Dejonghe \etal 1996).  Hence, it seems unescapable that \df s
depending {\it a priori} on three integrals should be available for
general models of elliptical galaxies.

In a stellar dynamical context, the problem of approximating the
 effective third integral is logically connected to the philosophy
 behind the construction method of dynamical models for galaxies:
 obviously, numerical models can do with numerical, or even implicit
 third integrals, while analytical models need analytical
 approximations.

Amongst the numerical modelling methods, those based on
Schwarzschild's approach (Schwarzschild 1979) are undoubtly the most
direct ones.  In a given potential, a library of orbits is computed,
while the time-averaged properties of the orbits are stored. The orbit
library is supposed to sample integral space.  A reproduction of the
observables is sought by combination of orbits, populated with a
non-negative number of stars ({\it e.g.} Rix \etal\ 1997; van der
Marel \etal\ 1998).  This powerful and very general technique allows
to construct general three-integral models, without any explicit
reference to the third integral, since in principle orbits can be
labelled using any phase-space point on them.  Spectral analysis
offers the possibility to simplify the expression for the orbits and
to reduce storage requirements for the orbits (Binney \& Spergel 1982;
Papaphilippou \& Laskar, 1996 \& 1998; Valluri \& Merritt 1998;
Carpintero \& Aguilar 1998).

However, for this type of modeling methods, the computational cost is
 relatively high. Moreover, smoothing is necessary because the
 singular boundaries of the orbital densities produce a fairly awkward
 sum of numerical functions, and the distribution function is a
 numerical function out of which the physical content may not be very
 easily extracted. The implicit reference to a third integral, which
 makes Schwarzschild methods so flexible, is at the same time a
 handicap for extracting information about the role played by this
 integral.

Analytical techniques would be more explicit.  But no expression for a
global third integral is known in general axisymmetric potentials --
except for \stapos\ ({\it cf.} de Zeeuw 1985; Dejonghe \& de Zeeuw
1988).  Therefore analytical approximations have been built using
perturbation methods.  Systems deviating moderately from spherical
symmetry have been considered, and an approximate integral derived,
which reduces to the total angular momentum in the spherical limit
(see Petrou 1983).  Gerhard \& Saha (1991) report on the use of three
different perturbation techniques  applied to a model that is
initially spherical. (1) Methods based on the KAM theory, which
have a validity that is often restricted to very small perturbations. 
In practice they find indeed that, whatever
their order, these methods fail to track changes in the phase-space
topology that occur when spherical symmetry is broken by a significant
amount. (2) The averaging method (see {\it e.g.} Verhulst 1979; de Zeeuw
\& Merritt 1983) in their example also fails to track the box--orbits 
($L_z=0$) that emerge in a
flattened system, and it gives at first-order only a rough
approximation to the third integral. (3) Finally, a resonant method
using Lie transforms does yield a good approximation to the third
integral.  It tracks the new orbital families emerging around
resonances, and may in principle be carried on to high order.  The
analytical expressions are however rather intricate, specially for
orders higher than 1.

The last technique has been used at first-order by Dehnen \& Gerhard
(1993) to construct three integral oblate models for the perturbed
isochrone sphere.  An application to the boxy E3-E4 galaxy NGC~1600
has recently been presented by Matthias \& Gerhard (1999).

However, one may need to model a galaxy whith a potential that is far
from any known integrable potential, so that global perturbation
techniques become ineffective.  An alternative approach is then to
derive a so--called `partial integral'. The idea is that a third
integral that can be easily determined for certain groups of orbits is
probably also a good approximation for similar orbits.
de Zeeuw, Evans \& Schwarzschild (1996) and Evans, H\"afner \& de
Zeeuw (1997) derived such a partial integral for thin and near-thin
tube orbits in scale-free potentials.

Yet another well known approach is the use of separable potentials.
One is led to this idea because motion around the origin can be
expanded in a Taylor series and, since the lowest-order non-zero terms
are quadratic, can be described as perturbed harmonic
motion. Moreover, van de Hulst (1962) has shown that motion around the
origin can be seen as (unperturbed) motion in a \stapo.  Basically,
one then uses a much enlarged set of reference integrable potentials
instead of merely quadratic ones.  The \stapo\ is found by fitting
term to term the Taylor series for the original potential (up to
quartic terms) to the expansion for the \sta\ potential.  Local fits
of similar nature have been performed by de Zeeuw \& Lynden-Bell
(1985) and Kent \& de Zeeuw (1991), who applied it to the solar
neighbourhood, by expanding the effective potential in the vicinity of
circular orbits.

However, such a local fitting does not work for all orbits. For
instance, it may fail for orbits with large radial extent. In many
cases, though, an orbit can still be approximated by motion in a
\stapo\, provided that the potential is chosen to be a good average
approximation to the true potential in the region covered by that
particular orbit (see Kent \& de Zeeuw 1991).  A report on the
calculation of such a good average and global approximation to the
Galactic potential, together with some indications for the numerical
implementation, can be found in Dejonghe \& de Zeeuw (1988) and
Batsleer \& Dejonghe (1994).

In this paper, we want to improve the generality of the application of
this family of potentials.  We propose to obtain a set of local
approximations to the potential using \stapos, each of which is an
average approximation to the original potential in some region.
Rather than expanding around an equilibrium point, which would make
the extent of the region where the approximation holds difficult to
handle, we partition integral space and fit the potential in the
corresponding regions of configuration space.  We use the QP method
(Dejonghe 1989) to adjust each local \stapo. Section 2 describes
this. Once the set of local \stapos\ is obtained, we proceed by
checking the quality of the orbits representation, and of the
approximation for an effective third integral, where it exists.
Obviously, we do not expect to describe the eventual stochastic motion
this way.  Moreover, we may not be able to reproduce all minor orbital
families.  These questions are also addressed in section 2.  In
section 3 the actual fitting procedure is explained. Its performance
is illustrated with three trial potentials, which correspond to mass
densities with {\it resp.} nearly ellipsoidal, boxy and disky
isophotes. These trial potentials are presented in section 4. The
results can be found in section 5. Section 6 gives a discussion on the
application of the method and the conclusions are given in section 7.

\section{The design of the set of St\"ackel potentials}
\subsection{The principle}
Orbits with a given set of integrals $(E,J)$, with $E$ the (positive)
binding energy and $J$ the component of the angular momentum along the
rotation axis, fill a certain volume $\cal S$ in space (a meridional
section of it is shown in figure 1b). The intersection of this volume
with a meridional plane is called the zero velocity curve (ZVC).  One
proves easily that orbits with a set of integrals $(E',J)$, with $E' >
E $ fill a volume that is completely embedded in the volume filled by
orbits with the set of integrals $(E,J)$.  Similarly, orbits with $E$
and $J'$ larger than $J$, have a ZVC that lies inside the ZVC defined
by $(E,J)$.
This means that all orbits in the shaded rectangle labeled $\cal R$ in
integral space (see figure 1a) will remain interior to the shaded region
indicated as $\cal S$ in Fig 1b. Any distribution 
function defined
over $\cal R$ in integral space can therefore be associated with a
potential that needs only be defined in $\cal S$.  Hence, a
subdivision of the $(E,J)$-plane into a number of rectangles $\cal R$,
is equivalent to subdividing space into a number of bounded domains
$\cal S$.  For each of these domains, a \stapo\ can be determined that
is locally a good approximation to the original potential.  If in the
end, the whole $(E,J)$-plane is covered by a set of rectangles ${\cal
R}_{(E,J)}$, also space will be covered by a set of domains ${\cal
S}_{(E,J)}$, and the original potential will be completely fit by a
set of locally fitted \stapos.

\subsection{Practically}

In practice, the specification of a set of rectangles $\cal R$ in integral
space is equivalent to the specification of a grid, each gridpoint of
which being the `upper left corner' of $\cal R$.  A grid in $E$ can
be constructed by dividing the system into equal mass shells and
computing the circular orbit energy for each shell radius. For every
value of $E$, we consider an equidistant grid in $J$, with 7 values
in the interval $]0,J_{\rm max}(E)[$, the upper bound corresponding to
the circular orbit with that energy (see also figure 2).

In St\"ackel potentials, orbits are determined by three integrals of
motion $(E,J,I_3)$.  Following van der Marel \etal\ (1998), the
different third integrals are parametrized with angles $\omega_i$,
that correspond to the angle between the horizontal axis and the
radius to the most external of the two points where the orbit
intersects the ZVC (see figure 2). The point of the ZVC where $I_3
={I_3}_{\rm max}$ is found. This is the point where the thin tube
orbit touches the ZVC. The coordinates $(R,z)$ of that point
determine the angle $\omega_{\rm max} = \pi -{\rm Arctan}
(z/(R-R_c(E))$.  The $\omega_i$ are chosen linearly between 0 and
$\omega_{\rm max}$, from which the values for $I_3(\omega_i)$ are
derived.

\begin{figure}
	\leavevmode
	\centerline{
	\epsfig{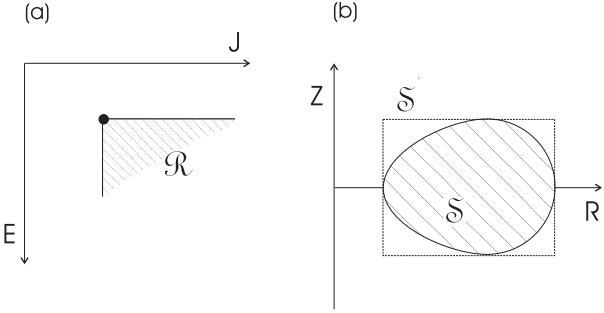}}
\caption{Panel (a): A rectangle $\cal R$ in integral space, of which
the \lq upper left corner \rq\ corresponds to a gridpoint
$(E,J)$. Panel (b): All orbits with $E'\ge E$ and $J'\ge J$ will remain
interior to the region $\cal S$. In practice, the local \stapo\ is
determined in the rectangle $\cal S'$. }
\label{cregrid}
\end{figure}

\begin{figure}
	\leavevmode
	\centerline{
	\epsfig{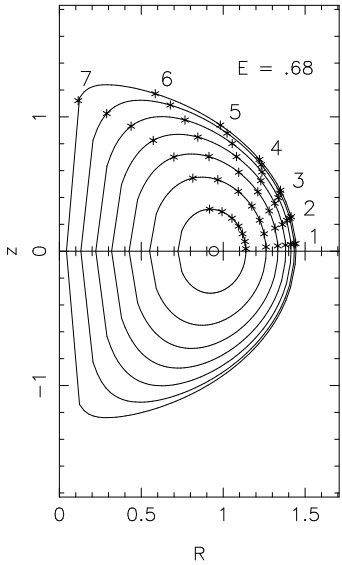}}
\caption{For one value of E in the grid : zero velocity curves
corresponding to 7 different values for $J$ and points used to
determine 7 values for $I_3$.}
\label{cregrid}
\end{figure}

As for the fit itself,  it suffices to perform the St\"ackel fit to the
potential associated with a rectangle ${\cal R}_{(E,J)}$ only in the
region interior to the velocity curve ${\cal S}_{(E,J)}$. In practice,
we fit within the smallest rectangle ${\cal S}'_{(E,J)}$ which encompasses
${\cal S}_{(E,J)}$.

The construction of a set of potentials that covers the complete
system potential is a multi-step process.  It is most efficient to
start with a global fit to the potential, i.e. fitting a \stapo\ in
the most extended domain $S'$, determined, e.g., by the data. This
domain defines the smallest value for $E$. The following steps improve
the fit to the potential by performing a number of local fits in
smaller domains.

The choice of domains where a fit will be performed is in principle
arbitrairy, and depends only on how well the \stapo\ has to
approximate the system potential (within the limitations of what a
regular potential can achieve).

\subsection{When is a fit considered to be a good fit?}

The fit is checked by comparing orbits in both potentials.  Orbits
start from the ZVC defined by $(E,J)$, at the point determined by
the value of $I_3$. The integration of orbits uses a fourth-order
Runge-Kutta scheme with variable time-steps, proportional to the
smallest of the radial and azimuthal periods, estimated respectively
as $T_R \sim R/V_R$ and $T_{\Omega} \sim 2 \pi R^2/J$.  This ensures
conservation of energy over $100\,T_{\Omega}$ with a precision better
than $ 10^{-6}$.

One can consider various criteria for the comparison:
\begin{itemize}
\item {\sl Surfaces of section:} The \stapo\ should generate
orbits that are very similar to those in the original potential.  A
first inspection can be done by comparing surfaces of section (SoS's) in
both potentials.  What we mainly expect to find this way is: 
\begin{itemize}
\item the proportion of non-regular orbits.  If there is no
additional integral of motion besides $E$ and $J$, the orbits should
fill uniformly the area bounded by the zero velocity curve on their
SoS (Richstone 1982). If there is a third effective integral
(effective meaning over the time the orbits are integrated), orbits
are regular, and appear as curves on the SoS.  Experience has taught
that in axisymmetric models, stochastic orbits exist only for very
small values of $J$.  In models with a core, as considered here, 
these orbits are    
due to the large excursions in $R$ and $z$ for small $J$, which
allow resonances between the two degrees of freedom (Merritt 1999).
\item the minor families of orbits which may be present around some
resonances in the original potential, that our \stapo\ does not
generate.  We call them unrecovered minor families.  The thin tube orbit
at fixed $(E,J)$ determines the main orbital family. This family
encircles the origin on the $(z,v_z)$ SoS, or a point $(R_{th},0)$ on
the $(R,v_R)$ SoS, where $R_{th}(E,J)$ is the point where the
thin-tube orbit of given $(E,J)$ intersects the equatorial plane.
Minor orbital families may be generated by other resonances than the
ones for small $J$, and will show up on a SoS as `islands' surrounding
other points.
\end{itemize}

\item {\sl Orbital densities:} 

Since the \stapo\ is aimed to be used for modeling, which is
essentially assembling orbits to fit a given density, a good
reproduction of the orbital densities is important.

These orbital densities $\nu(R,z;E,J,I_3)$ are functions of $(R,z)$
for each given orbit, defined for any set $(E,J,I_3)$.  The orbital
densities for both potentials can be compared by measuring the
fraction of the mass that is located in the same place in both
potentials. If the total mass of every orbit is normalised to 1,
and the cells surface is constant over the grid, the mass fraction
correctly located ($MC$) can be calculated as
\begin{equation}
MC=1- \delta M= 1-\sum_{k,\ell} |(\nu_{\rm or}(R_k,z_\ell) -
\nu_S(R_k,z_\ell)) | /2. 
\label{mc}
\end{equation}
In this, $\delta M$ is the mass fraction that is not located in the
same cell in both potentials, $\nu_{\rm or}$ stand for the original
density and $\nu_S$ stands for the density associated with the
\stapo. Double counting is avoided by dividing the sum by 2.

To calculate this, each orbit determined by a given set $(E,J,I_3)$, is
integrated over a large number of periods, and its orbital
density $\nu(R,z; E,J,I_3)$ (normalised to 1) is computed.
This is done on a rectangular grid covering
$[R_{min}(E,J),R_{max}(E,J)]\times[0,z_{max}(E,J)]$, by evaluating the
time fraction spent in each cell.

In principle the orbit should be integrated until the orbital density
reaches quasi-stationarity. This we may define by demanding that
during a given time interval the value $\nu_{kl}$ in each cell
$(R_k,z_l)$ has varied by less than some small fraction.
In practice, it takes an extremely long time to reach stationarity for
orbits close to a resonance $m:n$ with $m,n$ large, or having a
moderate degree of stochasticity -- these two cases being difficult to
distinguish ({\it cf} Binney \& Tremaine 1987, p.176).

Also modeling based on the Schwarzschild method has to deal with
stationarity (see {\it e.g.}  Schwarzschild 1993; Wozniak \& Pfenniger
1997).  Since modeling aims at constructing equilibrium models, only
orbits for which the orbital density reaches stationarity in limited
time 
({\it i.e.} less than a Hubble time)
 need in principle to be
taken into account. Accordingly, the non-stationary orbital densities
would not need to be precisely reproduced in the St\"ackel potential.

\item {\sl Conservation of $I_3$:}
If locally a \stapo\ is a good approximation, the St\"ackel third
integral $I_3$ should be approximately constant along the regular
orbits.  Since the goal is to provide a good approximation for an
effective third integral for the largest possible number of orbits,
the variation on $I_3$ should be made as small as possible.  If
possible, fits on smaller domains will be performed if this improves
the approximation for the integral.  The approximation is evaluated
through the maximal variation of $I_3$ along the orbit with respect to
its initial value, $\delta I_3/ I_3$.

On the other hand, one can argue that what really matters is a nominal
variation of $I_3$, $\delta I_3/{I_3}_{\rm max}(E,J)$ : for a given
$(E,J)$, how precise can our identification of orbits be when we
estimate the third integral of motion by $I_3$ within $[0,{I_3}_{\rm
max} (E,J)]$? A related question is how precise this identification
should be or, alternatively: how large is the tolerated variation on
$I_3$? 

The \stapos\ will be used as basis for a dynamical model. One of the
aims of these models is to reveal the role of the third
integral. Hence it is important to know that all structure caused by
the dependence on this third integral can be traced.  Consequently,
the \stapo\ should be precise enough, so that the uncertainty on the
resulting approximation for $I_3$ is not larger than the scale of the
smallest feature in $I_3$ of the \df. For example, in the limit where
the dynamical model would turn out to be essentially 2-integral, there
would be no need to set requirements on the constancy of $I_3$.

This makes determining the maximum tolerance for errors in the fit to
the potential an iterative process, which includes the modeling of
the galaxy. First, a fit to the potential has to be obtained, and a
model has to be constructed with that potential.  The results of the
model will then indicate whether the error on $I_3$
was small enough or not.

\item {\sl Topology of the orbits:}

As pointed out by Gerhard \& Saha (1991), approximate conservation of
$I_3$ along orbits alone does not ensure that $I_3$ provides a correct
labelling for orbits (see also de Zeeuw \& Merritt 1983).  One still
needs to check that the topology of orbital tori is similar to that of
constant $(E,J,I_3)$ surfaces: for then each orbit may be uniquely
associated to one value of $I_3$. This is done by comparing surfaces
of section for orbits in the original potential, and $I_3=cst$ curves
for given $(E,J)$.

\end{itemize}

\section{How to obtain a St\"ackel potential}

The fitting procedure is derived from a deprojection method for
triaxial systems presented in a paper by Mathieu \& Dejonghe (1996).
In that paper, a family of potential-density pairs is presented that
can be used as building blocks for triaxial mass models.  The spatial
mass density and the potential can be both expressed in terms of the
same basis functions.

Using this method, it would technically be possible to obtain a
St\"ackel approximation in two ways: (1) fitting the spatial mass
density (2) fitting the potential.

The first way is not ideally suited for performing local fits since
the calculation of a local value for the potential requires an
integration of the density over the entire volume of the galaxy.  In
the case of a fit in a restricted part of space, there is 
obviously  no control                      
on the behaviour of the mass density at larger distances,
 and this can
have serious implications for the potential obtained through
integration.
 The correspondence between the forces generated by the
original and the fitted potential, is likely to be a more useful
indication for the quality of the approximation than the
correspondence between the spatial densities. 
 For a global fit on the
other hand, it does make sense to construct a \stapo\ through a fit on
the spatial density.

The fitting procedure is based on a Quadratic Programming method.  The
aim is to find a linear combination of basis functions that provide a
good approximation to the original function.  When fitting the
potential, the basis functions are
\begin{equation}
F(\tau) = -{GM\over{(d+\tau^p)}^s}.
\label{ef}
\end{equation}
The basis functions and their coefficients are chosen by
minimizing the variable
\begin{equation}
\chi^2 = \sum_{k,\ell} w_{k\ell} \left[f_{\rm original}(r_k,z_\ell) -
f_{\rm calculated}(r_k,z_\ell)\right]^2,
\label{chike}
\end{equation}
with $k$ and $\ell$ an index covering the grid points in the domain $\cal
S'$. The weights $w_{k\ell}$ can be used to give different relative weights
to the points in the grid, $f$ stands for the function to fit.

In an ellipsoidal coordinate system $ -\gamma \le \nu
\le-\alpha\le\lambda$, with $\alpha$ and $\gamma$ negative, and the
focal distance $\Delta=\sqrt{|\alpha-\gamma|}$ (de Zeeuw 1985), a
\stapo\ can be written as
\begin{equation}
V(\lambda,\nu)=g_\lambda F(\lambda) + g_\nu F(\nu),
\end{equation}
with $F$ an arbitrary function (here corresponding to (\ref{ef})) and 
\begin{equation}
g_\lambda={{\lambda+\alpha}\over{\lambda-\nu}}.
\end{equation} 

We now present the relevant formulas from Mathieu \& Dejonghe (1996),
adapted for an axisymmetric system, that lead to an expression for the
density in terms of the basic functions.

The expression for the axisymmetric density is
\begin{equation}
\rho(\lambda,\nu) = g^2_\lambda \psi'(\lambda) + g^2_\nu \psi'(\nu)
+ 2 g_\lambda g_\nu \psi[\nu,\lambda],
\end{equation}
where $\psi[\nu,\lambda]$ is the first order divided difference of $\psi$:
\begin{equation}
\psi[\nu,\lambda] = \frac{\psi(\lambda)-\psi(\nu)}{\lambda - \nu},
\end{equation}
and $\psi'(\lambda)$ is the derivative $\psi[\lambda,\lambda]$.

The connection between $\psi(\tau)$ and $F(\tau)$ (like in (\ref{ef}))
is given by
\begin{equation}
2 \pi G \psi(\tau) = 2(\tau+\gamma) F'(\tau) - F(\tau) \nonumber
+ 2 {{\tau+\gamma}\over{\tau+\alpha}} 
\left[ F(\tau)-F(-\alpha)\right].
\label{gf}
\end{equation}

The density can be expressed in terms of the basic functions by means
of a third order divided difference:
\begin{equation}
\rho(\lambda,\nu)=H[\lambda,\nu,\lambda,\nu],
\end{equation}
with $H(\tau)$ defined as:
\begin{equation}
H(\tau)=(\tau+\alpha)^2 \psi(\tau),
\end{equation}
with $\psi(\tau)$ given in (\ref{gf}).

The divided difference of order $n-1$ of a function $G(\tau)$ is a
function of the divided difference of order $n-2$, and is given by
\begin{equation}
G[\tau_1,\ldots,\tau_n]={{G[\tau_1,\tau_3,\ldots,\tau_n]-G[\tau_2,\tau_3,\ldots,
\tau_n]}\over\tau_1-\tau_2}.
\end{equation}

\section{Three test potentials}

\begin{figure}
	\leavevmode
	\centerline{
	\epsfig{file=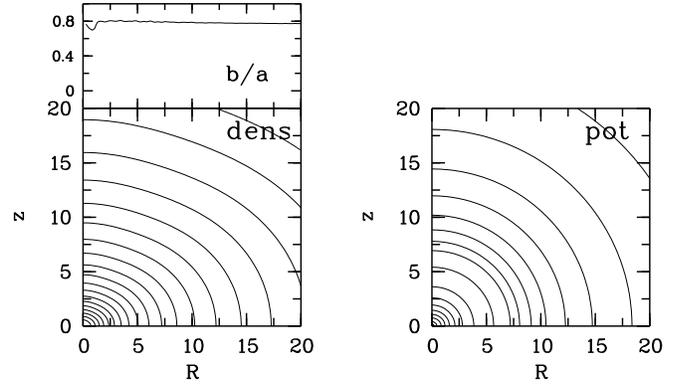, width=9cm, bbllx=30 ,bblly=210 ,bburx=470, 
bbury=470,clip=}}	
\caption{The spatial density, with its axis ratio, and potential
corresponding to $\Psi_{\rm ell}$, given by the sum of harmonics given
in equation (\ref{rhoharm}) and equation (\ref{potharm}), with
$\beta=-.3$.}
\label{harmdepo}
\end{figure}

We build a smooth flattened model by combining a few spherical
harmonics $Y^0_l(\theta)$ (Binney \& Tremaine 1987), as
follows:

\begin{eqnarray}
\rho(r,\theta) &=& C [\rho_0(r) Y^0_0(\theta,\phi) +
\beta \rho_2(r) Y^0_2(\theta,\phi)\nonumber\\
&+&{|\beta|\over 5} \rho_4(r) Y^0_4(\theta,\phi)],
\label{rhoharm}
\end{eqnarray}
\begin{eqnarray}
\rho_0(r)&=& {1\over (1+r^2)^2}\\
\rho_2(r)&=& {r^2\over (1+r^2)^3}\\
\rho_4(r)&=& \rho_2(r).
\end{eqnarray}
$\beta$ is a parameter which determines the flattening ($\beta<0$ for
oblate systems).  We set $C= 2/(\pi \sqrt{\pi})$ for a total mass
equal to $1$.  This simple choice for the model ensures $\rho \propto
r^{-4}$ and a fairly constant flattening. The resulting density for
$\beta= -0.3$ is shown in figure \ref{harmdepo}.  It is nearly
ellipsoidal, with axis ratio $b/a \simeq 0.8$.  We will in the
following refer to the corresponding potential as $\Psi_{\rm ell}$.

The corresponding potential is
\begin{eqnarray}
\psi_{ell}(r,\theta) &=& -4\pi G C [\psi_0(r) Y^0_0(\theta,\phi) +
\beta \psi_2(r) Y^0_2(\theta,\phi)\nonumber\\
&+&{|\beta|\over 5} \psi_4(r) Y^0_4(\theta,\phi)],
\label{potharm}
\end{eqnarray}
where the $\phi_l$ are related to the $\rho_l$ terms 
by equation (2-122) in Binney \& Tremaine (1987).

As explained in paragraph $\S$ 2, we use a grid in integral space that
defines the domains for the fit. The grid contains 16 points in E
(noted $E_i$), 7 points in $J$ (noted $J_i$) and 7 angles $\omega_i$.

For each of the domains, a fit can be produced, based on the $\chi^2$
of equation (\ref{chike}), if the focal distance $\Delta$ is set.  The
choice of $\Delta$ of the spheroidal coordinate system affects the
quality of the fit.  An average $\Delta$ is obtained by computing a
number of orbits in the original potential, and evaluating the focal
distance for which sections of the $\lambda=cst$ ellipses most closely
follow the inner and outer envelopes $R_{\rm min}(z)$, $R_{\rm
max}(z)$ of the orbits.  Of course, there is more variation in
$\Delta$ for the more extended domains $\cal S'$.

We started with a global fit for the potential $\Psi_{\rm ell}$, i.e. a
fit in the largest domain $\cal S'$ (extending to about 180 in $R$ and
in $z$), corresponding to the orbits that have ($E_{16}$,0).  We take
$J=0$ instead of $J_1$ in order to avoid that orbits with smaller $E$
should not be covered by the potential because their $J<(J_1)_{E_{16}}$.

This combination ($E_{16}$,0) is obviously the most unfavorable for
the fit, and the variations in $I_3$ are expected to set a standard
for the worst case: the maximum $(\delta I_3)_{\rm max}$ for all 7
orbits with different $\omega_i$ is taken as the maximum tolerable
variation. We demand that the variation of $I_3$ along the orbits with
lesser spatial extent (i.e. $E'>E_{16}$, $J'>0$ and the 7 $\omega_i$,
with $(E',J')$ on the grid in integral space) is not worse than
this. In this case, it turns out that the variation on $I_3$ is larger
for orbits with ($E_{15}$,$J_1$).  Thus the next fit is done in the
domain $\cal S'$ determined by ($E_{15}$,0).

After this fit, $(\delta I_3)_{\rm max}$ is the error on $I_3$ for the
orbits with ($E_{15}$,${J}_1$) and this new value is now used to
decide on the next fit.  In this way, fits are done in the domains
corresponding to ($E_{16}$,0), ($E_{15}$,0),
($E_{14}$,0), ($E_{11}$,0), ($E_2$,0) and
($E_1$,0).

\begin{figure}
	\leavevmode
	\centerline{
	\epsfig{file=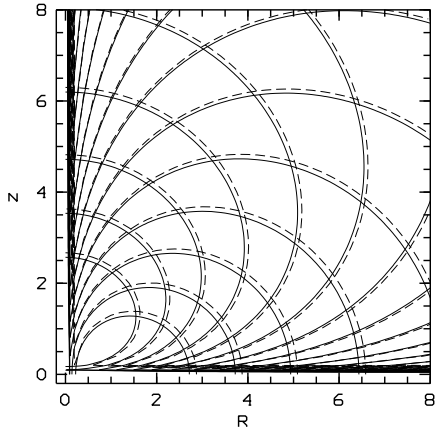, width=6cm, bbllx=45 ,bblly=55 ,bburx=340, 
bbury=340,angle=0,clip=}}
\caption{The contours for the two force components generated by
$\Psi_{\rm ell}$(full lines) and the \stapo\ for the domain $\cal
S'$(11,0)(dashed lines).  }
\label{forces}
\end{figure}

If the correspondence between the orbits in a fitted and original
potential happens to be very bad, it could be due to a poor
correspondence between the force components of both potentials.
Therefore, prior to the calculation of the orbits, it is useful to
check the behaviour of the force components in the fitted and the
original potential. Figure \ref{forces} shows the contours for the
force components generated by $\Psi_{\rm ell}$ (full lines) with the
\stapo\ for the domain ${\cal S}'(E_{11},0)$ (dashed lines) in
overlay. Since the force components in both potentials do not
appear too different, it is sensible to proceed with the comparison of
integrated orbits.

\begin{figure}
	\leavevmode
	\centerline{
	\epsfig{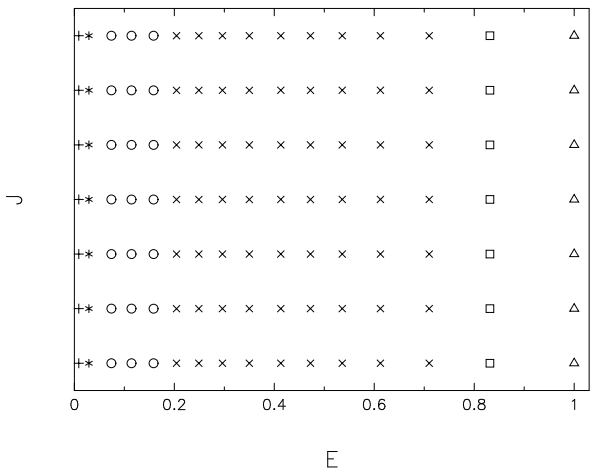}}
\caption{The $(E,J)$-grid for $\Psi_{\rm ell}$, with different
symbols for different potentials. We used a set of 6 \stapos to
approximate the original potential.}
\label{harmgrid}
\end{figure}

In figure \ref{harmgrid} the points in the $(E,J)$-grid for $\Psi_{\rm
 ell}$ are displayed as an example of how a set of potentials can be
 built; different symbols indicate the different potentials used. 
 As can be seen from the different symbols, some potentials of the set
 are used for a small number of orbits, while others are used for a
 large number of orbits. The fact that the potential fitted in the
 largest domain is only used for orbits with only one value for $E$
 (indicated by the pluses) suggests that this method can really
 improve the approximation.  Also the orbits that remain close to the
 centre seem to require a potential fitted in a limited part of
 space.

\begin{figure}
	\leavevmode
	\centerline{
	\epsfig{file=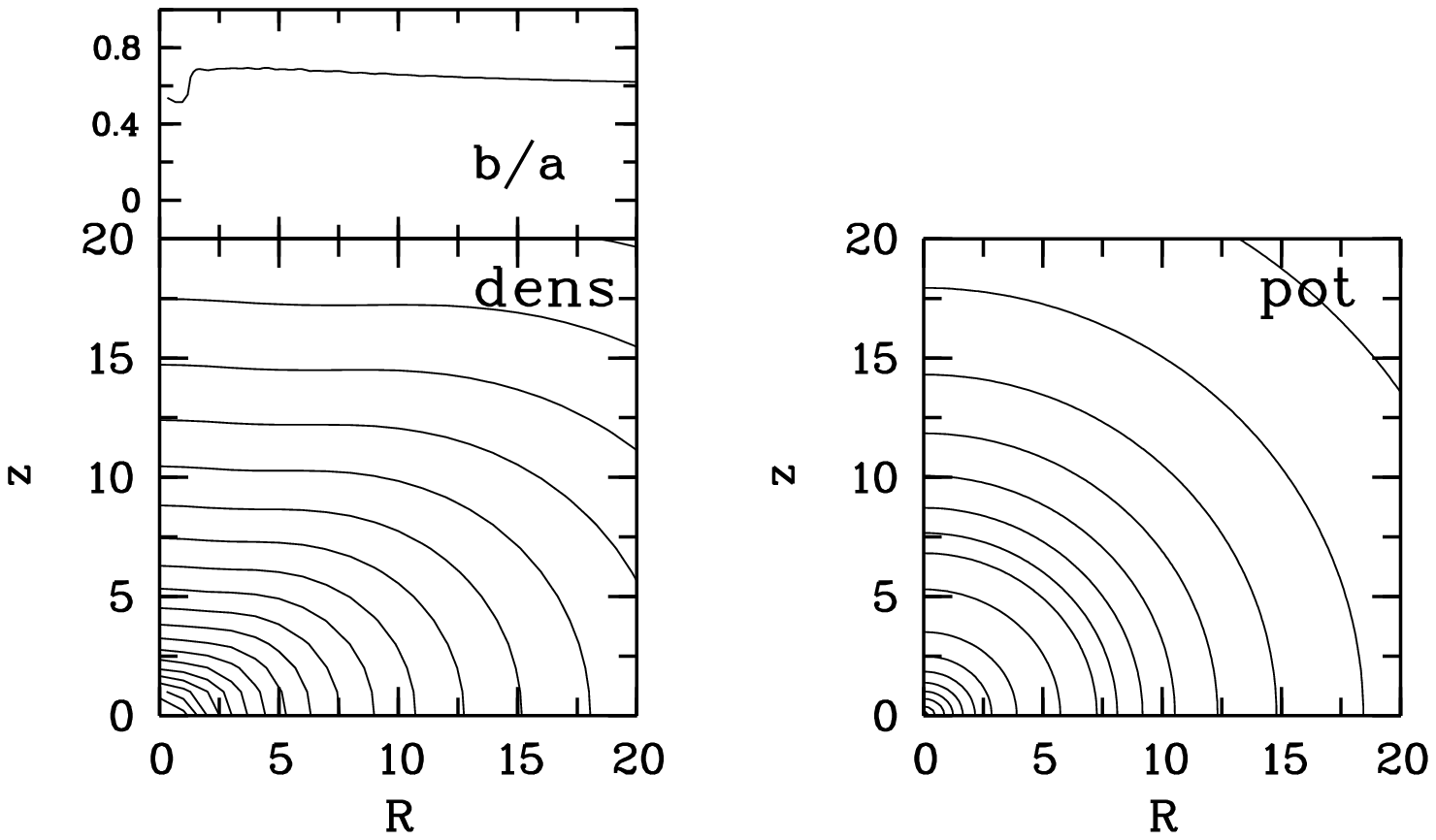, width=9cm, bbllx=30 ,bblly=210 
,bburx=470, bbury=470,clip=}}	
\caption{The spatial density, its axial ration and the potential
corresponding to $\Psi_{\rm Box}$, given by the sum of harmonics given
in equation (\ref{rhoharm}) and equation (\ref{potharm}), with
$\beta=-.5$.}
\label{boxdepo}
\end{figure}

As a second trial case, a boxy density, with potential $\Psi_{\rm
box}$, is obtained from a combination of harmonics as described in
(\ref{rhoharm}) and (\ref{potharm}) with $\beta= -0.5$, and has an
axis ratio $b/a \simeq 0.6$.  The contours are shown in figure
\ref{boxdepo}.  We also used a $16_E\times 7_J \times 7_{I_3}$ grid
and approximated $\Psi_{\rm box}$ with a set of 8 \stapos.

\begin{figure}
	\leavevmode
	\centerline{
	\epsfig{file=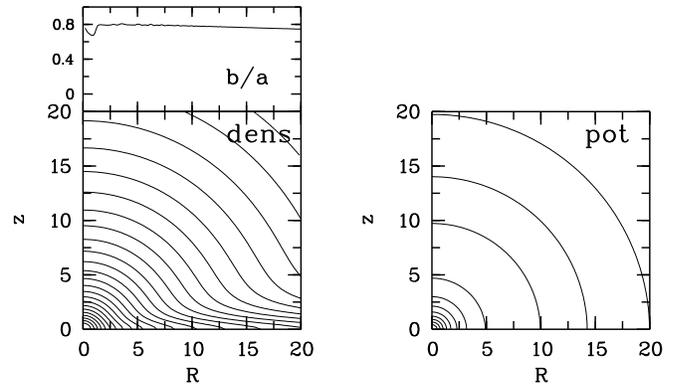, width=9cm, bbllx=30 ,bblly=210 ,bburx=470, 
bbury=470,clip=}}
\caption{The spatial density, its axial ratio and the potential for a 
Miyamoto-Nagai
potential with $b/a\sim 0.7$. The contours for the spatial density
are strongly disky.}
\label{mn2depo}
\end{figure}

Also a Miyamoto-Nagai (MN) potential with intermediate flattening,
$b/a\sim 0.7$, is taken as trial potential. 
MN models exhibit very strong diskiness in the density contours, as
can be seen in figure \ref{mn2depo}. This means that we are
considering a difficult case, that is actually on the verge of being
unrealistic for elliptical galaxies.

The grid has the same dimensions as the one described in the previous
section. The MN potential is approximated with 8 potentials, the set
of \stapos is constructed following the same strategy.

\section{Presentation of the results}
\subsection{Surfaces of section}

Typical surfaces of section display ($R,v_R$) at $z=0$, for the
orbits having $v_z > 0$ when they cross the equatorial plane or
display ($z,v_z$) for $v_R =0$, for those orbits having $v_R >0$ just
after crossing.

The SoS's immediately tell us that for the three potentials, all
orbits of our phase-space grid have an effective third integral.
Stochasticity, if present, was mild and would have required very long
integrations to be visible on the SoS's.  We did not carry out such
integrations, because the SoS's were only meant as a preliminary
check. In practice, anyway, mildly stochastic orbits behave very much
like regular orbits, and we can hope to approximate them by an orbit
in a \stapo.

For $\psi_{\rm ell}$ and $\psi_{\rm box}$, all orbits in our library have
similar SoS's in both potentials.  There is no evidence for
stochasticity, nor for unrecovered minor families.  For the MN
potential, all orbits also appear as regular, but we found a few
unrecovered minor orbital families.

Most of the orbits trapped around resonant tube orbits -- what we
called `minor orbital families' in \S 2.3 -- were present in the local
\stapo.

\begin{figure}
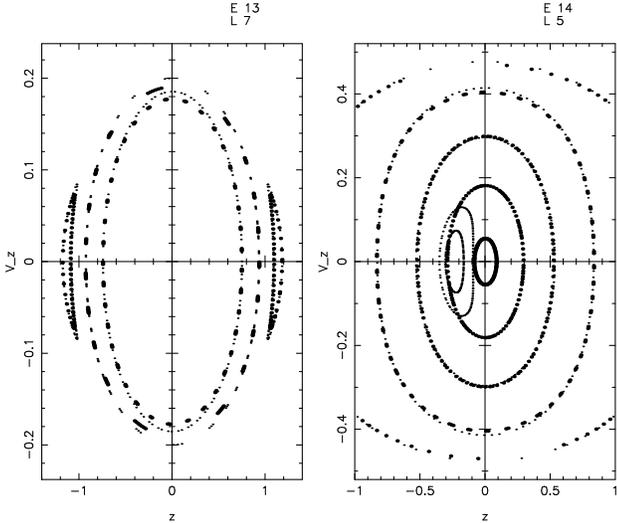

	\leavevmode
	\centerline{\hbox{
	\epsfig{file=sos_137.ps, width=7cm, bbllx=30 ,bblly=31 ,bburx=580, 
bbury=350,angle=-90,clip=}	
\epsfig{file=sos_154.ps, width=7cm, bbllx=30 ,bblly=31 ,bburx=580, 
bbury=350,angle=-90,clip=}}}
\caption{Surfaces of section for typical orbits in the MN potential (small dots) 
and the \stapo (large dots).}
\label{SoS}
\end{figure}

In figure \ref{SoS} SoS's of a few orbits are displayed for the MN
potential (small dots) and the \stapo\ (large dots).  Both panels show
a `minor resonance' evidenced by small `islands'. In the left panel,
showing orbits with large $I_3$, the orbits trapped around minor
resonances are well reproduced within the \stapo.  The right panel
displays orbits with smaller $I_3$. Two orbits are trapped around a
1:1 resonance which does not exist in the \stapo. They are examples of
unrecovered minor orbital families.  They are always found for small values of
$I_3$. Therefore, they correspond to orbits remaining close to the
equatorial plane, where the MN potential is very much distorted by a
strong diskiness.

Even for such extreme `diskiness', these orbits only represent $< 1\%$
of the orbit library for MN. These obviously are orbits that we will
not be able to take into account when building a model that uses
\stapo s. Although the unrecovered minor families are only a small
fraction of the entire orbital library, we will exclude them from the
statistics of the other checks discussed hereafter. As such, we are
not taking those orbital families into account that no one expects to
be approximated by a \stapo.

\subsection{Orbital weights}

For every orbit, the conserved mass fraction $MC$ is calculated
following equation (\ref{mc}), using a $20 \times 20$ grid in $(R,z)$.
We integrate each orbit for at least $50\,T_{\Omega}$, at most $200\,
T_\Omega$.  We check every $50\,T_{\Omega}$ whether quasi-stationarity
has been reached.  Quasi-stationarity is assumed when for every cell
the orbital density has varied by less than $1\%$ in $50 \,
T_{\Omega}$.

\begin{figure}
	\leavevmode
	\centerline{
	\epsfig{file=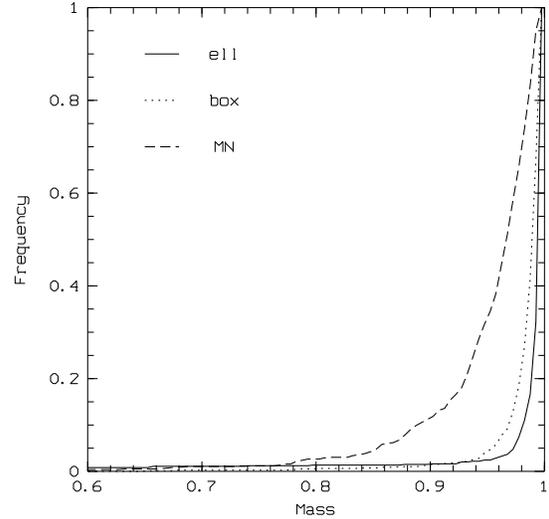, width=7cm, bbllx=51 ,bblly=45 ,bburx=530, 
bbury=540,angle=-90,clip=}}
\caption{Cumulative distribution for the fraction of mass 'correctly
located' in the orbital densities. The full line is for $\Psi_{\rm
ell}$, the dotted line is for $\Psi_{\rm box}$ and the dashed line
for MN.}
\label{cumul_harm}
\end{figure}

The results are presented in figure \ref{cumul_harm} which gives the
cumulative distribution for $MC$, the fraction of mass 'correctly
located' in the orbital densities.  A few orbital densities for MN
have $MC< 60\%$, these correspond to minor resonances that
were not fitted by our \stapos.  Excluding those, the average $MC$ is
$99\%$ for $\Psi_{\rm ell}$, $98\%$ for $\Psi_{\rm box}$ and  $95\%$ for MN.

\subsection{How constant is the St\"ackel $I_3$?}
\begin{figure*}
	\leavevmode
	\centerline{
	\epsfig{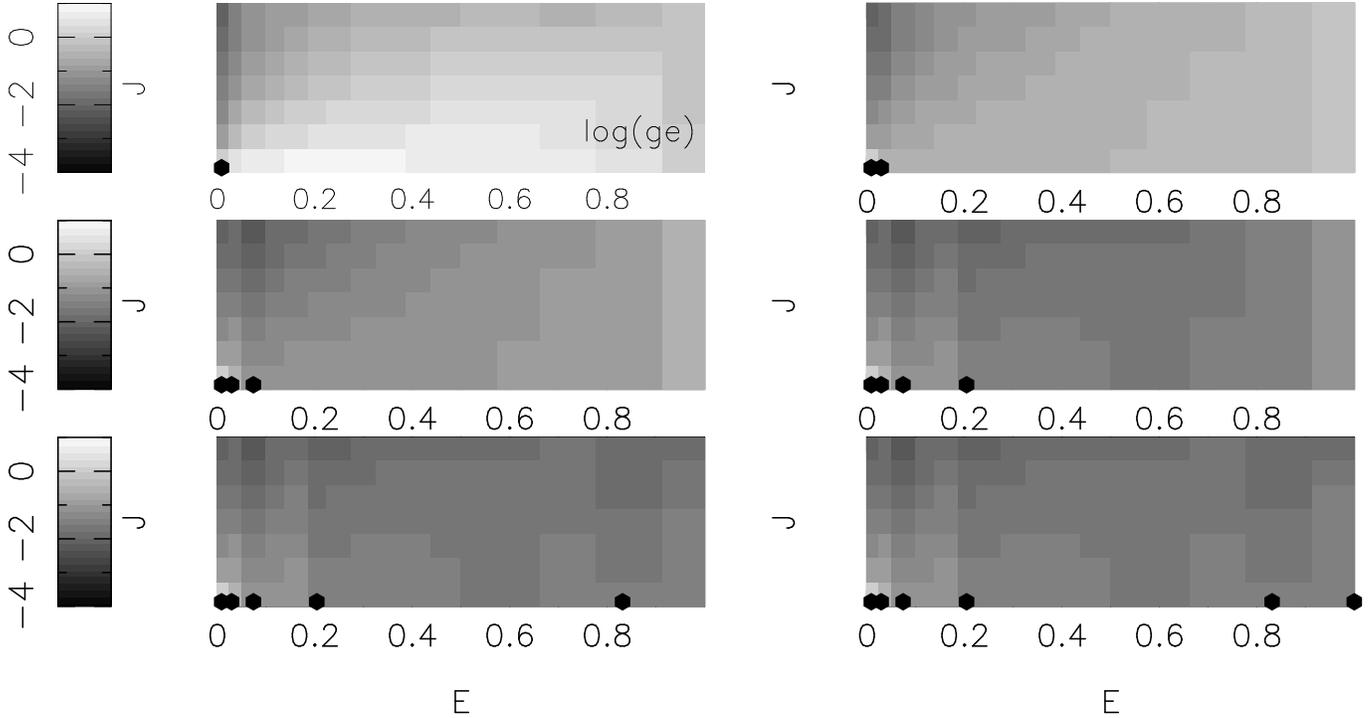}
}
\caption{The evolution in $log(\delta^r)$ as more potentials are
added to the set of \stapos\ approximation $\Psi_{\rm ell}$. The black
dots indicate the $E$ of the domain where the potentials were fitted
($J=0$ for all six potentials).  Dark ({\it resp.} light) gray shades
represent small ({\it resp.} large) variations on $I_3$.}
\label{patchharm}
\end{figure*}

For every potential, the conservation of $I_3$ along the orbits is
estimated by computing the relative variation $\delta^r \equiv \delta
I_3(E,J,I_3)/I_3(E,J,I_3)$.  At given $(E,J)$, the maximum of this
quantity is derived among orbits with different $I_3$: $\delta^r_{\rm
max} \equiv \max_{i=1,7} [ \delta I_3(E,J,I_{3,i})/I_3(E,J,I_{3,i})
]$.

Figure \ref{patchharm} shows how $\log(\delta^r_{\rm max}) $ is
affected as more potentials are added to the set approximating
$\psi_{\rm ell}$.

The black dots represent the $E$ ($J=0$) of the domain $\cal S'$ where
the potentials are fitted.  Dark gray shades represent small
variations on $I_3$, light gray shades represent larger
variations. For each new local potential that is considered, a number
of previously void cells of the $(E,J)$ grid are filled and a number
of values are replaced by new ones.  It is clear that adding new
potentials to the set improves the conservation of the third integral
along the orbits, the difference between the global fit (upper left
panel of figure \ref{patchharm}) and the results of the complete set
(lower right panel) is remarkable.

\begin{figure}
	\leavevmode
	\centerline{
	\epsfig{file=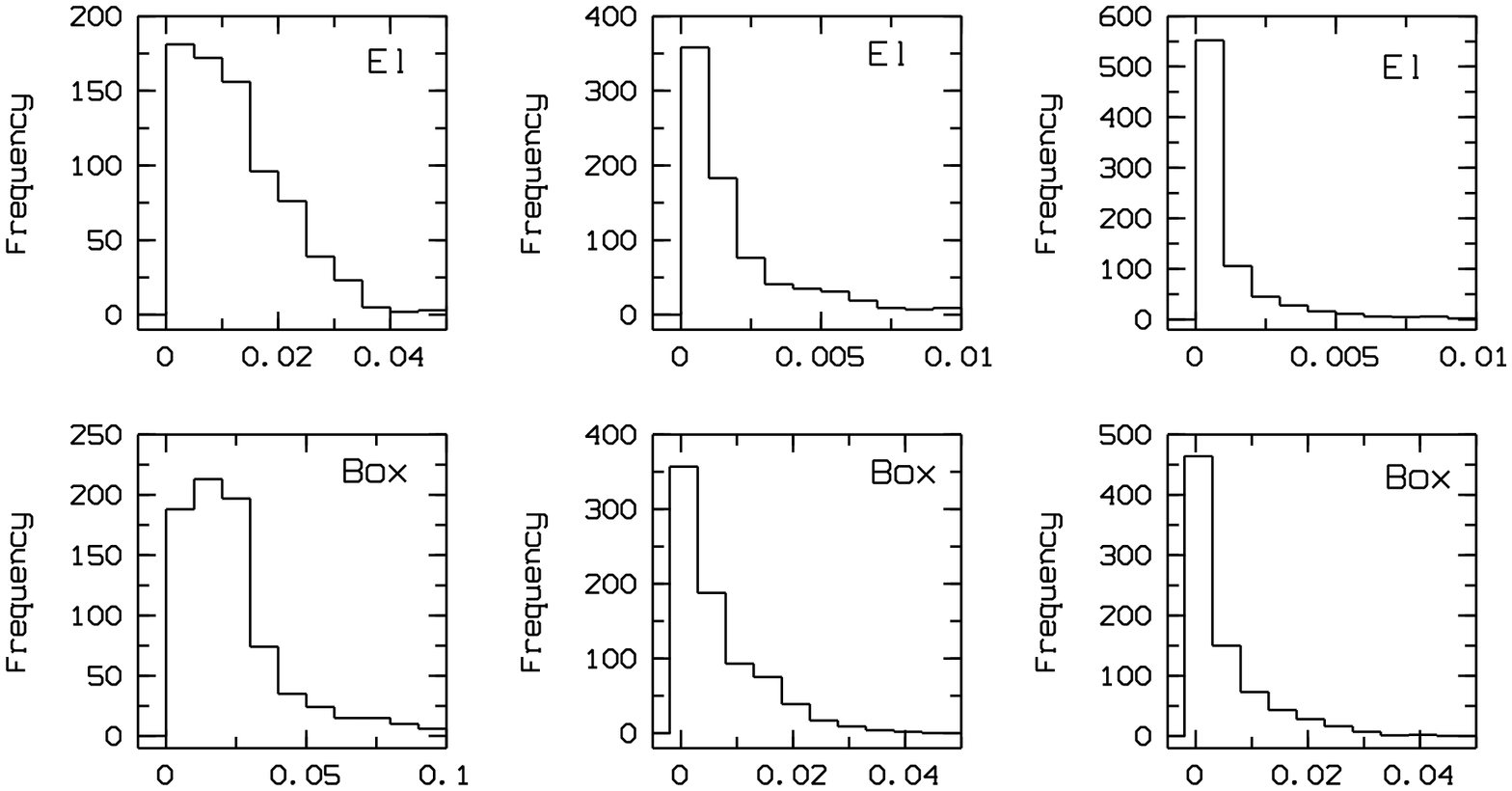,height=9cm, bbllx=50 ,bblly=50 ,bburx=370, 
bbury=670,angle=-90,clip=}}	
	\centerline{
	\epsfig{file=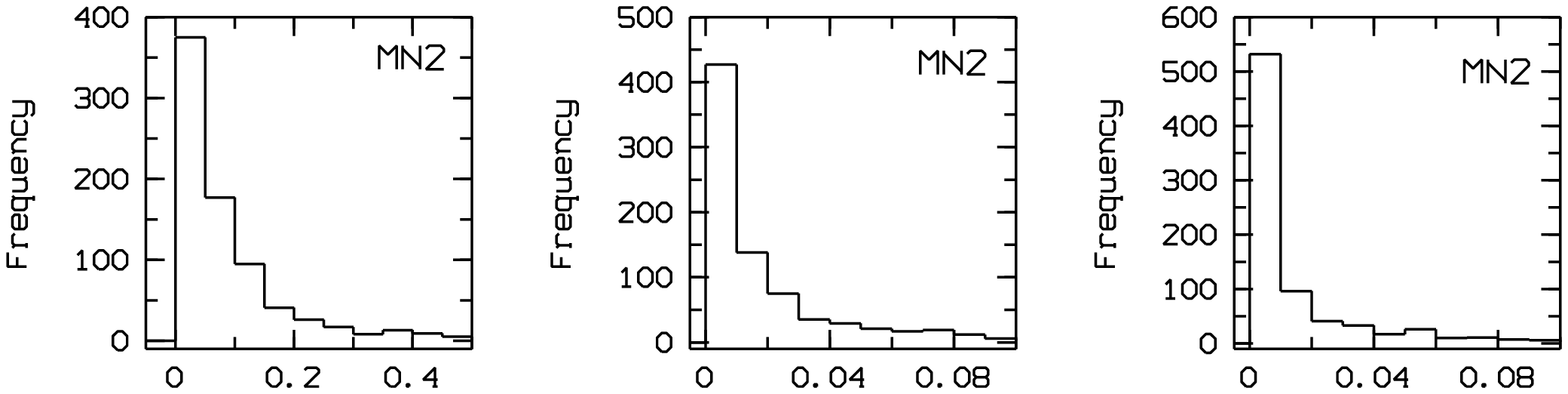,height=9cm, bbllx=50 ,bblly=50 ,bburx=220, 
bbury=670,angle=-90,clip=}}
\caption{Histograms of $\delta^r$ in the left panels, $\delta^n$
(i.e. $\delta I_3$ normalised to ${I_3}_{\rm max}(E,J)$) in the middle
panels and of $\delta I_3$ normalised to ${I_3}_{\rm max}(E)$ in the
right panel.}
\label{consi3}
\end{figure}

In the left panels of figure \ref{consi3} we present histograms of
$\delta^r$ for from top to bottom: $\psi_{\rm ell}$, $\psi_{\rm box}$ 
and MN. For the MN potential, the resonances absent from the
St\"ackel approximation are not considered, a complete orbital library
contains 784 orbits.

The average for $\delta^r$ is $2\%$ for $\psi_{\rm ell}$, while $75\%$
of the orbits have $\delta^r<1.7\%$. For $\psi_{\rm box}$ the average
for $\delta^r$ is $\sim 2.6 \%$ and $75\%$ of the orbits have
$\delta^r<2.7\%$. For MN, the average value is $\sim 10 \%$, and
$75\%$ of the orbits have $\delta^r<12\%$.  The variation of
$\delta^r$ along each orbit is of the order of the fitting error on
the derivatives of the potential.

In the middle panels of figure \ref{consi3} the results in terms of a
nominal variation $\delta^n \equiv [\delta I_3(E,J,I_3)/{I_3}_{\rm
max}(E,J)] $ are shown.  For the nearly ellipsoidal potential
$\Psi_{\rm ell}$, $75\%$ of the orbits have $\delta^n<.3\%$, with an
average value of $\sim .2\%$. For $\Psi_{\rm box}$ the average is
$\sim .7\%$ while $75\%$ of the orbits have $\delta^n<1\%$.  For MN,
the average value is $\sim 1.7\%$ and $75\%$ of the orbits have a
nominal error on $I_3$ less than $ 2.3\%$.

Dehnen \& Gerhard (1993) used a similar indicator, but they normalised
$\delta I_3$ to ${I_3}_{\rm max}(E)$ , which gives smaller values for
the nominal variation, as can be seen in the right panels of figure
\ref{consi3}. The average for respectively $\Psi_{\rm ell}$,
$\Psi_{\rm box}$ and MN is $.1\%$, $.5\%$ and $1.3\%$.

\subsection{Topology of constant $I_3$ varieties}
\begin{figure}
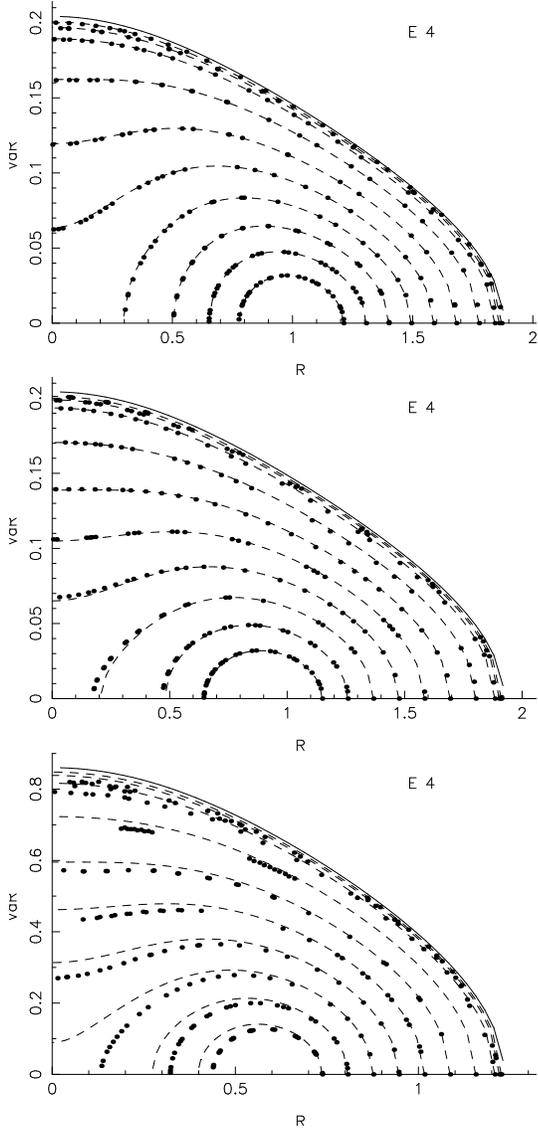

	\leavevmode
	\centerline{
	\epsfig{file=harmp3.ps, height=7cm, bbllx=120 ,bblly=45 
,bburx=570, bbury=680,angle=-90,clip=}}
	\centerline{
	\epsfig{file=harmp5.ps, height=7cm, bbllx=120 ,bblly=45 
,bburx=570, bbury=680,angle=-90,clip=}}
	\centerline{
	\epsfig{file=mn2.ps, height=7cm, bbllx=120 ,bblly=45 ,bburx=570, 
bbury=680,angle=-90,clip=}}

\caption{Comparison of SoS's and $I_3$ level curves at given
($E_4,J=0$), for $\Psi_{\rm ell}$ (upper panel), $\Psi_{\rm box}$
(middle panel), and $MN$ (lower panel).  Dots: surfaces of section
$(R,\dot R)$ at $z=0$ with $\dot z > 0$.  Lines: the intersection of
$I_3=cst$ surfaces with the $z=0$ plane.}

\label{top}
\end{figure}

We plot on the same graph (figure \ref{top}): (i) the $(R,\dot{R})$
SoS of orbits as they cross the $z=0$ plane (with $\dot{z} >0$) in the
original potential; and (ii) the invariant curves defined by $I_3=
cst$ (at $z=0$).  This is shown here for some intermediate value of
$E$ ($E_4$ of our grid) and $J=0$.  Radial orbits are in principle
difficult to map, because of the transition between loop and box
orbits (see Gerhard \& Saha 1991).  They are however very well
described here, both for $\Psi_{\rm ell}$ and $\Psi_{\rm box}$. The
agreement is somewhat poorer in the case of MN for the
transition region between box and loop orbits; however, the topology
is still correctly described, in the sense that we can establish a
one-to-one correspondence between the two sets of curves.  This shows
that $I_3$ may be used to label the orbits.

\section{Discussion}
For the construction of dynamical models, the use of \stapos\ yields a
number of advantages, amongst them (1) the property that the density
and dynamical moments can be calculated analytically and (2) the
absence of regularisation problems. Finally, once the dynamical model
has been completed, the distribution function is known in an easy to use
and analytical form.  Though these models require a special form of the
potential, they offer a great flexibility during the further modeling
process.  In a \stapo\, there is a function of 1 variable that can be
freely chosen. This freedom is advantageous, and can, of course, be
exploited at the fullest when performing a fit to the galaxy
potential.

Analogous to local \stapo s , dynamical models built within
this approximation will yield local distribution functions,
simply because we have adapted the coordinate
systems in each domain $\cal S$  where a local \stapo is constructed. The
use of a QP-method (Dejonghe, 1989) for dynamical modeling gives a
large freedom for the basis functions. In this case, we can use basis
functions for the distribution function that contribute only in
limited parts of integral space. In practice, these limited parts will
correspond to the rectangles $\cal R$ from the local potentials.

The fact that the space of \stapos\ has measure zero in the space of
all axisymmetric potentials, raises the obvious concern whether
\stapos\ are suitable for the representation of any galaxy potential.

The scientific objections cluster around the following arguments: 

(1) `What about central cusps?'  Indeed, central density cusps
generally cannot be generated by a \sta\ potential in an ellipsoidal
coordinate system. However, Sridhar \& Touma (1997) showed that a
\stapo\ in parabolic coordinates can create a central cusp.  Whether
these potentials can be used in the scheme we propose here remains to
be investigated.  On the other hand, it may well be possible that
there is no point in providing a good approximation to the third
integral in regions where the cuspiness of the density becomes
important.  Indeed, it is likely that the central density cusps are
related to the presence of a central massive object (see {\it e.g.}
review by Kormendy \& Richstone 1995; Richstone \etal\ 1998).
This may cause a higher fraction of ergodic orbits over a Hubble time.
Such orbits will fill regions of phase--space where the distribution
function depends essentially on the energy, or, for less ergodic
regions, also on angular momentum ({\it cf} Merritt 1999).  The
fraction of regular to ergodic motion in a real galaxy stands as a
major unknown. The most popular scenario at the moment is that
ergodicity has gradually been introduced in the system, as orbits that
pass close to the central mass concentration were perturbed by it
(Gerhard \& Binney 1985; Valluri \& Merritt 1998).  One may then
suppose that the original orbital tori which have not been disrupted
still underlie most of the features in phase-space-- with essentially
homogeneous 6-D density between them.

(2) `Does a St\"ackel approximation allow an accurate fit to a typical
  galactic potential?'  As was already noticed for some time, the main
  orbit families found by numerical integration in general triaxial
  potentials are present in a \stapo\ (Schwarzschild 1979; de Zeeuw
  1985), but obviously there is no place in an integrable potential
  for smaller orbital families nor stochastic orbits.  These minor
  orbital families appear to occupy only a small volume fraction of
  phase-space, as long as figure rotation is unimportant (Gerhard,
  1985; Binney, 1987) In practice, St\"ackel potentials
  do turn out to provide reasonably good global fits for systems
  without central mass concentration (Dejonghe \etal\ 1996) or for
  regions beyond the influence of the central mass concentration
  (Emsellem \etal\ 1999). It is true that the use of one single
  \stapo\ assumes a single confocal system where the streamlines of
  the mean stellar motion coincide with the coordinate lines of the
  ellipsoidal coordinate system. Hence, the construction of such a
  single \stapo\ inevitably involves some sort of averaging (de Zeeuw
  \& Lynden-Bell 1985; Dejonghe \& de Zeeuw 1988), and is not always
  considered to be a sufficiently general approach (at least in
  triaxial cases, Binney 1987; Merritt \& Fridman 1996). However,
  the use of confocal streamlines seems to be a valid assumption for
  individual orbits (Anderson \& Statler 1998).  Moreover, studying
  logarithmic potentials, those authors find that the mean velocities
  can be well fitted using local coordinate systems with confocal
  coordinate lines, with the focal distances taken within a fairly
  narrow range.  Our approach allows to carry this idea one step
  further, on the level of the distribution function.

\section{Conclusions}
In this paper, we present an extension of the available approximations
for the effective third integral in axisymmetric systems, obtained by
using a set of \stapos\ as representation for a galaxy potential,
instead of one single \stapo.  We have studied the feasibility and
effectiveness of this method.

The creation of a set of local \stapos\ is done through a fit on the
system potential.  There is a large freedom in the choice of the
domains where the local fits are done and the composition of the set
of \stapos.  The set of approximating local potentials can be extended
until the desired precision is obtained.

We have tested the method on three potentials: (A) a harmonical
potential ($\Psi_{\rm ell}$) that behaves very smoothly with a nearly
ellipsoidal density; (B) a harmonical potential ($\Psi_{\rm box}$)
with a boxy density; and (C) a Miyamoto-Nagai model (MN) with a
density that has an exaggerated disky structure.

As one may expect, the model with smallest diskiness/boxiness is easiest
to approximate using sets of St\"ackel potentials.
But what really motivates this study is to check the
estimate made for the effective third integral, 
if we approximate it by the third integral $I_3$ that the
\stapos\ define.

The quality of the approximation is checked in several ways: (1)
surfaces of section, that reveal possible resonances and irregular
orbits, (2) conservation of orbital weights, which is important for
the reconstruction of the spatial density, (3) conservation of $I_3$
along the orbits, in order to validate the labeling of orbits, (4)
topology of the orbital space.  The conservation of orbital weights
and the conservation of $I_3$ are the criteria that can be best
expressed numerically.

According to the expectations, the approximation of $\Psi_{\rm ell}$
has orbits that are very similar to the orbits in the original
potential, as can be judged from the surfaces of section and the
topology.  The results for the conservation of orbital weights (an
average of $99\%$ for $\Psi_{\rm ell}$) and $I_3$ ( $\delta^r \equiv
\delta I_3(E,J,I_3)/I_3(E,J,I_3) \sim 2 \%$ and $\delta^n \equiv
[\delta I_3(E,J,I_3)/{I_3}_{\rm max}(E,J)] \sim.2\%$ for $\Psi_{\rm
ell}$) confirm the quality of the approximation.

The method also proves to be successful for $\Psi_{\rm box}$, with an
average of $98\%$ for the conservation of orbital weights, $\delta^r
\sim 2.6\%$ and $\delta^n \sim.7 \%$. The SoS's and the topology of
the orbits also confirms this good result.

The success of the method on the Miyamoto-Nagai potential is somewhat
less, because of the strong diskiness. Still, using a small set of
\stapos, we are able to reproduce most orbits with satisfactory
accuracy, except for a few resonances. The resonances that are not
reproduced by the set of \stapos, all have small values of $I_3$,
i.e. the orbits lie in the region where the diskiness is important.
Leaving this resonances out of consideration, the orbitals weights
seem to be well conserved (on average $95\%$ for MN) and the topology
of the orbits is well reproduced. Also for these potentials, the
St\"ackel $I_3$ can be used as label for the orbits ($\delta^r\sim10\%$
and $\delta^n\sim1.7\%$). The potentials of observed elliptical galaxies
are generally rounder than the strongly disky Miyamoto-Nagai models.

Given the positive results found for the three rather different models
considered, it seems to be possible, using a reasonable number of
local \stapos, to provide good approximations for a third integral,
suitable for labeling orbits in dynamical models.

With respect to existing approximations for a third integral in
axisymmetric systems, the advantage of this new one is mainly that,
while it can be envisaged for application to systems with arbitrary
flattening, it also yields a simple analytic expression for the
approximate third integral locally.  This method will be fully
exploited if it is used to build, for roughly axisymmetric galaxies,
explicit \df s that depend on three integrals.

{}

\end{document}